\input harvmac.tex

\input epsf
\input tables.tex
\noblackbox
\nref\one{P. Candelas and X. de la Ossa, Nucl. Phys. {\bf B341} (1990) 383.}
\nref\two{I. Klebanov and M. Strassler, JHEP 0008 (2000) 052, 
hep-th/0007191; C. Vafa, 
J. Math. Phys. {\bf 42} (2001) 2798, hep-th/0008142.}
\nref\three{M. Bershadsky, V. Sadov, and C. Vafa, Nucl. Phys. {\bf B463} 
(1996) 420, hep-th/9511222.} 
\nref\five{J. Blum, JHEP 0001 (2000) 006, hep-th/9907101.}
\nref\six{B. Greene, A. Shapere, C. Vafa, and S.-T. Yau, Nucl. Phys.
{\bf B337} (1990) 1.}
\nref\seven{G. Papadopoulos and P. Townsend, Phys. Lett. {\bf B380} (1996) 
273, hep-th/9603087.}
\nref\eight{A. Tseytlin, Phys. Lett. {\bf B381} (1996) 73, hep-th/9604035.}
\nref\nine{A. Fayyazuddin and D. Smith, JHEP 9904 (1999) 030, hep-th/9902210.}
\nref\ten{H. Cho, M. Emam, D. Kastor, and J. Traschen, Phys. Rev. {\bf D63} 
(2001) 064003, hep-th/0009062.}
\nref\eleven{P. Green and T. H{\"u}bsch, Int. J. Mod. Phys. {\bf A9} (1994) 3203,
hep-th/9303158.}
\nref\cdm{G. Cardoso, B. de Wit, and T. Mohaupt, Class. Quant. Grav. 
{\bf 17} (2000) 1007, hep-th/9910179.}
\nref\dm{K. Dasgupta and S. Mukhi, Phys. Lett. {\bf B398} (1997) 285,
hep-th/9612188.}
\nref\kmvhi{S. Katz, P. Mayr, and C. Vafa, Adv. Theor. Math. Phys. 
{\bf 1} (1998) 
53, hep-th/9706110; K. Hori and C. Vafa, hep-th/0002222; K. Hori, A. Iqbal,
and C. Vafa, hep-th/0005247.}
\nref\s{A. Strominger, Nucl. Phys. {\bf B451} (1995) 96, hep-th/9504090.}

\def\tx{\vbox{\sl\centerline{Physics Department}% 
\centerline{University of Texas at Austin}%
\centerline{Austin, TX 78712 USA}}}

\Title{\vbox{\baselineskip12pt
\hbox{UTTG-03-02}\hbox{hep-th/0207016}}}
{\vbox{\centerline{Triple Intersections and Geometric Transitions}}}

{\bigskip
\centerline{Julie D. Blum}
\bigskip
\tx

\bigskip
\medskip
\centerline{\bf Abstract}
The local neighborhood of a triple intersection of fivebranes 
in type IIA string theory is shown
to be equivalent to type IIB string theory on a noncompact Calabi-Yau
fourfold.  The phases and the effective theory of the intersection are
analyzed in detail.
}

\Date{7/02}

\newsec{Introduction}

Many supersymmetric gauge theories are obtained as a sector of
string theory in a particular geometric background.  Varying the
moduli of the string geometry may allow passage between different
backgrounds and open a window onto the strong coupling regions of
the gauge theory.  Some of the phases are better described by
open strings while others are described by closed strings with
nonzero fluxes of background fields.

The conifold is a noncompact Calabi-Yau (CY) threefold described
geometrically as a cone over an ${\bf S}^2\times {\bf S}^3$ base.
The transitions between the geometries where an ${\bf S}^2$ is
blown up (${\cal O}(-1)+{\cal O}(-1)\rightarrow {\bf P}^1$)
and an ${\bf S}^3$ is blown up ($T^* {\bf S}^3$) have been studied
\one. Recently,
the transitions between open strings in one of the conifold geometries
and fluxes in the other have been studied as a means to describe
phases of $N=1$ gauge theories \two .  The conifold has a T-dual
description as a pair of orthogonally intersecting Neveu-Schwarz (NS)
fivebranes on ${\bf R}^3$ \three .  Our aim in the present paper is to
understand similarly the triple intersection of NS fivebranes on a string.
Our motivation was to provide evidence for conjectural bound states
associated to this intersection \five .  After T-duality the triple
intersection has a geometric description, and one can describe the
phases through geometric transitions.  Work is in progress to determine
the spectrum of BPS states and possibly the metric.

In section two we will discuss the supergravity solution for triple 
intersections of fivebranes in eleven dimensional supergravity.
The solution will be reduced to ten dimensions along one of the
transverse directions and shown to be equivalent to the stringy
cosmic string construction of \six .  Finally, a T-duality along
three directions (generally, mirror symmetry for a CY threefold)
transforms the solution to a CY fourfold.  In section
three we present a linear sigma model describing the local
neighborhood of the singularity  of the CY fourfold.
The model turns out to be ${\cal O}(-1,-1)+{\cal O}(-1,-1)
\rightarrow {\bf P}^1\times{\bf P}^1$.
We show that this model has flops 
to isomorphic phases as well as
transitions to CY fourfolds with various other cycles.  In section four we
show that the perturbative modes of the supersymmetric 
spacetime two-dimensional
theory of a Dirichlet fivebrane in the ${\bf P}^1\times{\bf P}^1$ 
phase match with those in the 
dual fivebrane picture.  The dual fivebrane is a manifold partly 
constructed from a fivebrane.  Anomaly cancellation for the 
fivebrane guarantees
consistency in the CY fourfold framework.  We conjecture that a fivebrane 
instanton is the mechanism for resolving the singularity.  A local mirror
for the CY fourfold is given.  
The other phases are 
compared and also seem to agree.
The analogy with the conifold leads to various conjectures discussed in
this section.  Our results indicate that chiral anomalies only occur in 
the ${\bf P}^1\times{\bf P}^1$ phase where the fivebranes are unseparated and
that the theories are effectively four-dimensional in the other phases.

\newsec{Supergravity Solution of Intersecting Fivebranes}

We start with the following configuration of fivebranes preserving
one-eighth of the supersymmetry in eleven dimensional supergravity.

\vskip 20pt
\begintable

$0$|$1$|$2$|$3$|$4$|$5$|$6$|$7$|$8$|$9$|$10$\cr
x|x|x|x|x|x|||||\cr
x|x|x|x|||x|x|||\cr
x|x|||x|x|x|x|||
\endtable

\noindent The metric and equations that result from solving the supersymmetry 
constraints for this configuration have been determined in \seven\eight\nine
\ten .  The metric can be written in the following form.
\eqn\metriceleven{\eqalign{ds_{11}^2&=H^{-1/3}(-{dx^0}^2 +{dx^1}^2)+2H^{-1/3}
g_{m{\bar n}}
dz^m dz^{\bar n}+H^{2/3}dx^{\alpha} dx^{\alpha}\cr
g&=Hf{\bar f}\cr}}

\noindent In the above $f$ is an arbitrary holomorphic 
function of the $z^m$, $z^1=x^2+ix^3$, $z^2=x^4+ix^5$, $z^3=x^6+ix^7$,
and $\alpha\in\{8,9,10\}$.  Note that $g$ is the determinant of the $3\times 3$
matrix $g_{m{\bar n}}$.  The threefold metric  $g_{m{\bar n}}$ is required by
supersymmetry to be Kahler.  The field strength of the three-form gauge
field takes the following form.
\eqn\fieldstr{\eqalign{F_{{\bar n}\alpha\beta\gamma}&= -i\partial_{\bar n}
\ln H \epsilon_{\alpha\beta\gamma}\cr
F_{m\alpha\beta\gamma}&= i\partial_m
\ln H \epsilon_{\alpha\beta\gamma}\cr
F_{m{\bar n}\beta\gamma}&=-i\partial_{\alpha}g_{m{\bar n}}H^{-2/3}
\epsilon^{\alpha}\, _{\beta\gamma}\cr}}

\noindent The equation of motion for the gauge field yields 
\eqn\motion{2\partial_m\partial_{\bar n} H+\partial_{\alpha}^2 g_{m{\bar n}}=
J_{m\bar n}^{source}}

\noindent where $J_{m\bar n}^{source}$ is the magnetic source for the fivebranes.  These
equations are interesting to analyze directly in eleven dimensional 
supergravity since they resemble a generalized system of monopoles.
For our present purposes, however, we need to reduce to type IIA string
theory by taking $x^{10}$ to be a hidden compact dimension.  The metric
becomes
\eqn\metricten{ds_{10}^2=-{dx^0}^2+{dx^1}^2+g_{m{\bar n}}dz^m dz^{\bar n}+
H(dz d{\bar z})}
where $z=x^8+ix^9$, $H=e^{2\phi}={g\over f{\bar f}}$ ($\phi$ is the dilaton).
The field strengths are
\eqn\fieldstrten{\eqalign{H_{mz{\bar z}}&={-1\over 2}\partial_m H=-\partial_m
g_{z{\bar z}}\cr
H_{{\bar n}z{\bar z}}&={1\over 2}\partial_{\bar n} H=\partial_{\bar n}
g_{z{\bar z}}\cr
H_{m {\bar n}z}&=-\partial_z g_{m{\bar n}}\cr
H_{m{\bar n}{\bar z}}&=\partial_{\bar z}g_{m{\bar n}},\cr}}
and the equation of motion is 
\eqn\motionten{\partial_m\partial_{\bar n} H+2\partial_z
\partial_{\bar z} g_{m{\bar n}}=
J_{m\bar n}^{source}.}
Defining the Kahler form $J=i g_{m{\bar n}}dz^m \wedge dz^{\bar n}$ leads
to the following equation for the Kahler parameters, $\rho=B+iJ$.
\eqn\kah{\partial_{\bar z}\rho=0}
This is precisely the mirror of the stringy cosmic string construction
where one has $\partial_{\bar z}\tau=0$ for the complex structure
deformation parameters
$\tau$ of a Kahler threefold.  Note that the complex fourfold total space
fibered over the $z$ plane with fiber the Kahler threefold $X_3$ is not
Kahler since $\partial_z J_{m{\bar n}}\neq 0$.  The would be Kahler form
$J^{total}$ of the fourfold satisfies the equation 
\eqn\source{(d^2 J^{total})_{z{\bar z}m{\bar n}}={i\over 2}
J_{m{\bar n}}^{source}}
since the fivebranes
contribute to the stress-energy tensor.  The equations of
motion can be solved by taking $g_{m{\bar n}}$ to be a CY metric dependent
on $z$ and 
piecing together neighborhoods covering the CY in which
$\partial_m\partial_{\bar n}H= 0$.  By replacing $X_3$ by its mirror
$\tilde X_3$ in type IIB, we obtain the stringy cosmic string
construction of generally noncompact CY fourfolds \eleven .  There are
constraints on the number of strings, and it remains unclear whether 
a Ricci flat fourfold can be constructed for all $\tilde X_3$ (in particular
ones where the complex structure degenerates at infinity). 

In this paper we will restrict the analysis to the simplest case of
a ${\bf T}^2\times {\bf T}^2\times {\bf T}^2$ fibration over the 
$z$ plane in which each
fiber degenerates at one or more points in the $z$ plane.  
Performing a T-duality along one cycle of a torus exchanges the
Kahler parameter, $\rho=B+i\sqrt G$, of the torus ($B$ is the NS two-form,
$G$ is the determinant of the ${\bf T}^2$ metric) with the complex structure
parameter $\tau$ of the torus.  T-dualizing on one cycle of each torus
yields a ${\bf T}^2\times {\bf T}^2\times {\bf T}^2$ fibration 
over the $z$ plane in which 
the $\tau$ parameters are holomorphic functions of $z$, and the $\rho$
parameters are independent of $z$.  To avoid questions of global
consistency due to orbifold points on the tori, we will study the
local neighborhood of the degeneration where the three fibers are
noncompact.  The ${\bf T}^2\times 
{\bf T}^2$
case corresponds to the conifold while a ${\bf T}^2$ fibration that degenerates
at one point corresponds to a Kaluza-Klein fivebrane 
and is nonsingular.  There is additional freedom in the
above solution to add various fluxes of background fields.  The 
supergravity analysis is not usually valid in the limit where the
number of fivebranes is small and the sources are delta functions, but
this analysis has provided insight into the nature of the singularity.
In the next 
section we will determine a noncompact CY fourfold corresponding to
this degeneration.

\newsec{Geometry of the Triple Intersection}

The equations describing the local neighborhood of the degenerate fibers 
in the triple ${\bf T}^2$ fibration can be written in the form
\eqn\degfiber{z=a_1 a_2=b_1 b_2=c_1 c_2 .}
The singularity can be expressed as the intersection of two quadrics in
${\bf C}^6$.
\eqn\quadricsingone{a_1 a_2-b_1 b_2=0}
\eqn\quadricsingtwo{b_1 b_2-c_1 c_2=0}
Each equation by itself describes a conifold so the two equations
represent the intersection of two conifolds.  Doing a small resolution
of each conifold (blowing up a ${\bf P}^1$) yields four phases--each
conifold has two flops.
There is one large resolution describing the deformations of intersecting
${\bf S}^3$'s.  Additionally, there are six phases in which one
conifold is blown up and the other is deformed.  We will present each of
these phases in detail and show that there are transitions in which
the holomorphic four-form is preserved.

\subsec{The Small Resolution}

The small resolution is described by the following four equations for 
eight complex variables.
\eqn\small{\eqalign{a_1&=z_1 b_1\cr b_1&=z_2 c_1}\qquad
\eqalign{b_2&=z_1 a_2\cr c_2&=z_2 b_2\cr}}
The flops are generated by $a_1\leftrightarrow a_2$ and
$c_1\leftrightarrow c_2$.  Actually,
there are many possible deformations of the small resolution when one
takes into account deformations of  equation \quadricsingtwo .  Fixing
\quadricsingone\ by a linear transformation, there is still an $SO(4,{\bf C})$
invariance.  Equation \quadricsingtwo\ then has $21-12-1=8$ complex
deformations where we have subtracted an overall scale.  This counting 
agrees with the fivebrane side.  All of these deformations are
nonnormalizable since the fivebrane is noncompact.  We conjecture that
there are no deformations of the small resolution preserving
supersymmetry.
In section four we will find evidence for this conjecture by obtaining
a fivebrane manifold which looks very dissimilar to this small
resolution but produces an equivalent theory.

This resolution is also described by a gauged linear sigma model with two
$U(1)$'s and the following charges for six chiral fields $x_i$ under the
two $U(1)$'s.
\eqn\charge{\eqalign{l_1&=(1,1,-1,-1,0,0)\cr l_2&=(0,0,-1,-1,1,1)\cr}}
We can partly visualize the resolution in the following ``toric'' 
diagram that projects four dimensions onto a plane.  Generically,
the diagram depicts a ${\bf T}^4$ fibration that shrinks to a ${\bf T}^3$ on
three-dimensional boundaries enclosed by three lines, a ${\bf T}^2$ on
two-plane boundaries, an ${\bf S}^1$ on lines in the toric base, and
a point at the intersection of four lines.

%Diagram 1
\centerline{\epsfxsize=0.5\hsize\epsfbox{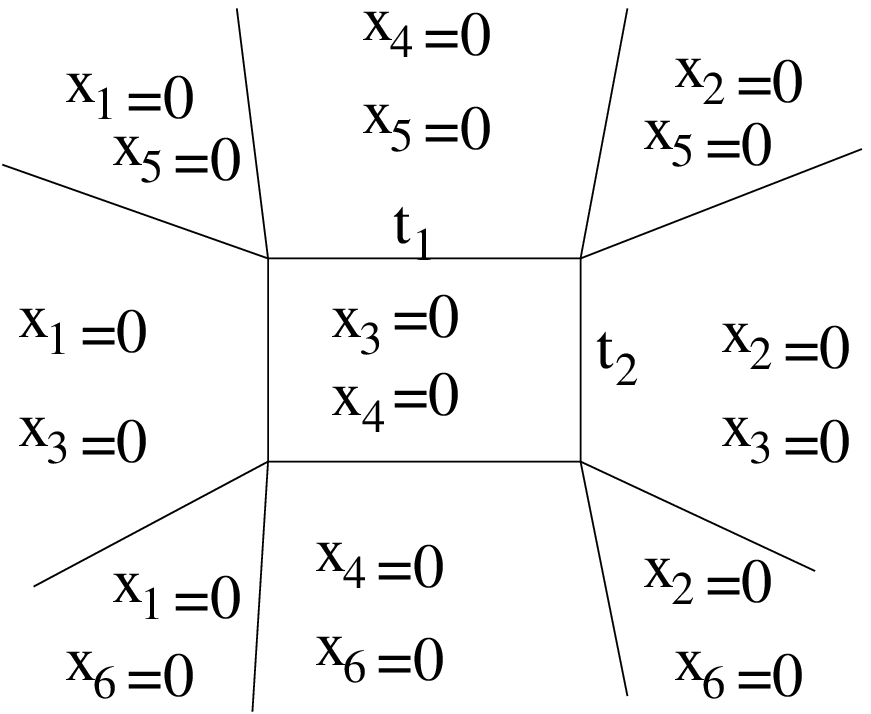}}
\bigskip
\centerline{\vbox{\noindent{\bf Diag. 1.}
Toric Diagram of the Small Resolution}}

\noindent Here, $t_1$ and $t_2$
are the Kahler parameters for the two ${\bf P}^1$'s.  Note that there
are other two-planes not labeled in the diagram.

\subsec{The Large Resolution}

To describe the large resolution we pick a nongeneric deformation 
of the moduli space 
that can be understood fairly easily and then argue that if we 
deform along a path that does not pass through singular regions, topological
features of the manifold should remain invariant.  Because the CY
fourfold is noncompact, most of the parameters of the deformation are not
normalizable.  
The nongeneric deformation that we choose is
\eqn\nongen{\eqalign{a_1 a_2-b_1 b_2&=\mu\cr b_1 b_2-c_1 c_2&=\nu\cr}}
with $\mu$, $\nu$ real and $\mu>\nu>0$.  The base of this manifold can
be depicted schematically in the following diagram as a four-torus 
fibration.

\vskip .5cm
%Diagram 2
\centerline{\epsfxsize=0.5\hsize\epsfbox{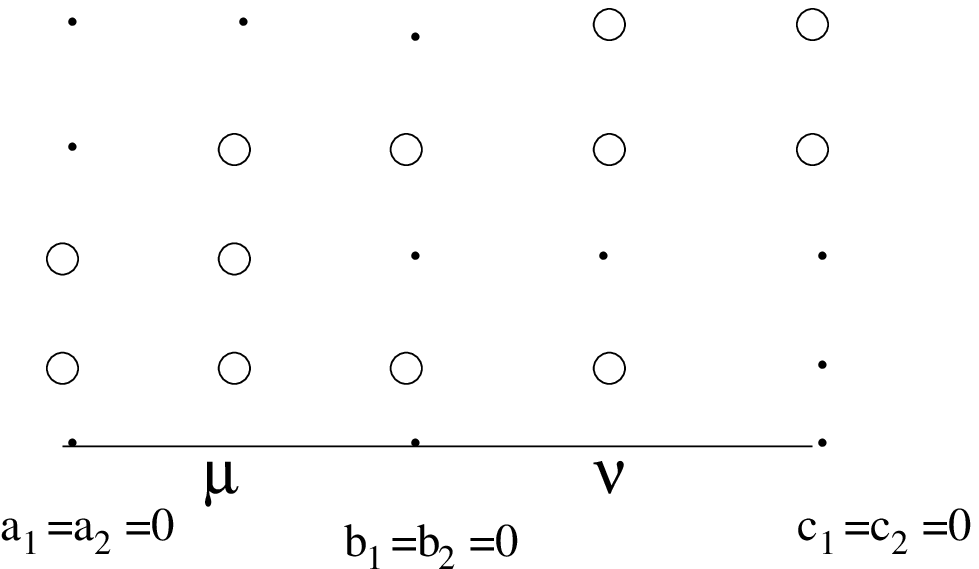}}
\bigskip
\centerline{\vbox{\noindent{\bf Diag. 2.}
Base of the Large Resolution}}

One can describe the space as $T^*{\bf S}^1$ fibered over $T^*{\bf S}^3$
where the $T^*{\bf S}^1$ degenerates away from the base ${\bf S}^3$.
The two independent ${\bf S}^3$'s are noncontractible
with sizes determined by $\mu$ and $\nu$.  The manifold is simply 
connected.  Each ${\bf S}^3$ intersects the other two ${\bf S}^3$'s
along a ${\bf T}^2$.  There is a third ${\bf S}^3$ with size parameter
$\mu+\nu$ that intersects $a_1=a_2=0$ and $c_1=c_2=0$ but not $b_1=b_2=0$
or anywhere else on the above diagram.  The ${\bf S}^3$'s are not isolated
since the $T^*{\bf S}^1$ direction is a zero mode direction.  
There are also two independent
noncontractible four-cycles with topology ${\bf S}^3\times{\bf S}^1$.
The ${\bf S}^3\times{\bf S}^1$ is homologous to a four-cycle where the 
${\bf S}^1$ shrinks over a portion of the ${\bf S}^3$ at loci where
the $T^*{\bf S}^1$
fiber degenerates, but there
is no flat direction preserving the area of the ${\bf S}^3$ and
connecting the two four-cycles.  
Another picture of the base as a three-torus fibration 
is presented below with $\mu$ real and
$\nu\rightarrow -i\nu$.  (Angles are not drawn accurately.)  In the diagram 
$k={{\mu-i\nu}\over\sqrt{\mu^2+\nu^2}}$.
The three edges are noncontractible four-cycles of topology 
${\bf S}^3\times{\bf S}^1$ bounding an open five-chain.

\vskip .5cm
%Diagram 3
\centerline{\epsfxsize=0.5\hsize\epsfbox{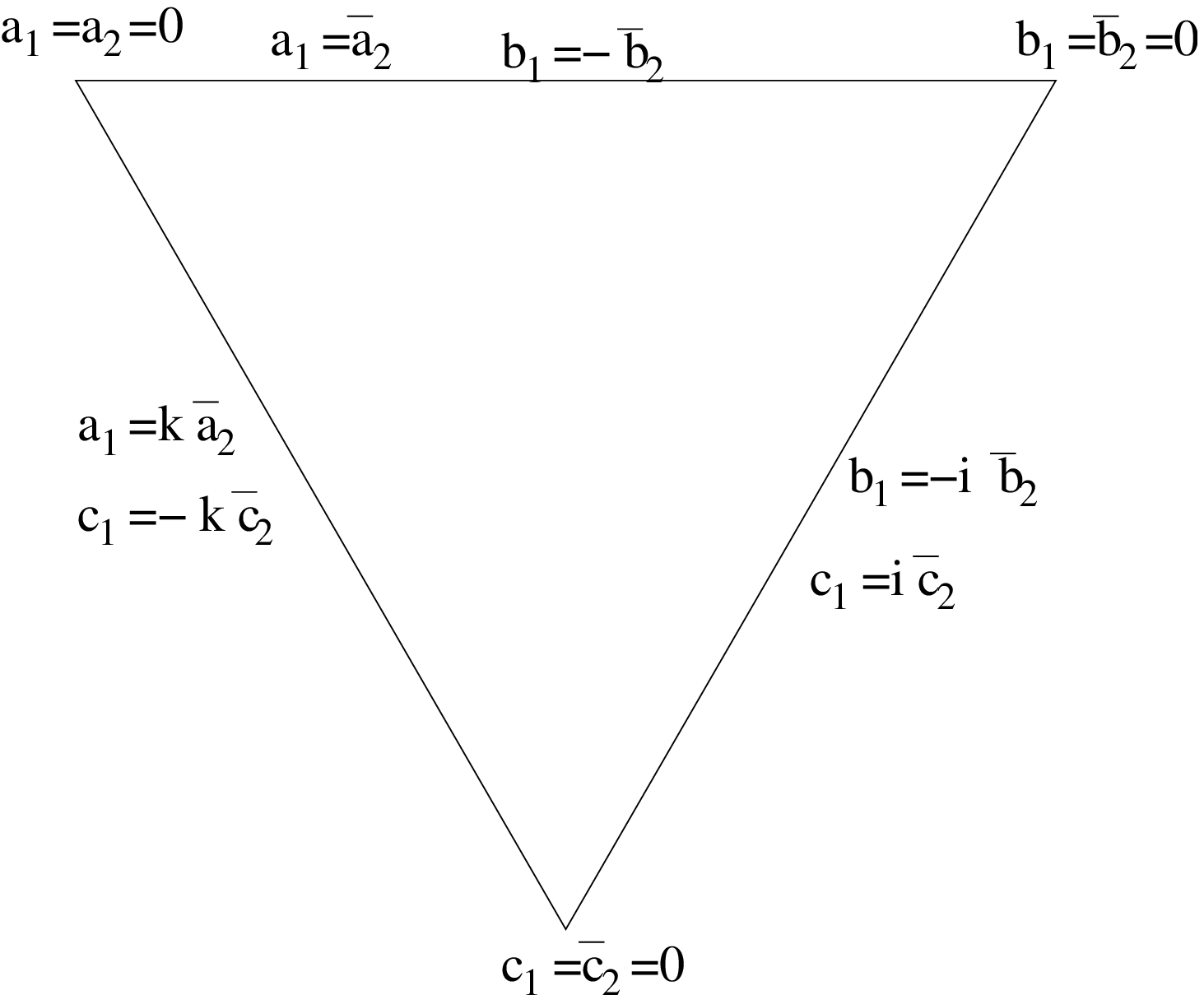}}
\bigskip
\centerline{\vbox{\noindent{\bf Diag. 3.}
Base of the Large Resolution ($\mu$ real, -i$\nu$ imaginary)}}

To deform the moduli space in a generic way we choose coordinates 
such that the form of the first equation is preserved.  Writing the second
equation as 
\eqn\deform{x_i G_{ij}x_j=\nu}
with $x_1=a_1$, $x_2=a_2$, $x_3=b_1$, $x_4=b_2$, $x_5=c_1$, $x_6=c_2$,
and $\nu$ complex, one can analyze the condition of transversality.
For every symmetric, nonzero matrix $G$, there will be a possible
nontransverse intersection of the hypersurfaces if
$det(G_{ij}-a\delta_{ij}\epsilon_j )=0$ where $\epsilon_1=\epsilon_2= 
-\epsilon_3=-\epsilon_4=1$ and $\epsilon_5=\epsilon_6=0$.
Nontransversality implies that $a={\nu\over\mu}$ for one of the roots $a$.
Obviously, this condition will not be a generic one.  One can show that
transversality implies that the holomorphic four-form has no zeroes or 
poles.  The deformation should preserve the structure of three 
three-cycles that are topologically ${\bf S}^3$ (two are independent)
intersecting as described above.  In terms of the original fivebranes,
this deformation should correspond to varying the holomorphic four cycle
on which one of the fivebranes is wrapped without compromising
transversality.  From this point of view there are nine complex
parameters determining the deformation.  At a first glance there are
twenty-one complex parameters deforming the second equation where we
fix $\nu$.  However, there is an $SO(4,{\bf C})$ group of symmetries
preserving the undeformed equation so that we are left with nine parameters in
agreement with the fivebrane count.  There are also the modes
$\mu$ and $\nu$ which control the sizes of the ${\bf S}^3$'s and will
turn out not to be normalizable.

\subsec{The Mixed Resolution}

We are calling the mixed resolution the CY fourfold obtained by doing
a small resolution on one of the two conifolds and a large resolution
on the other.  There are six isomorphic phases and we will describe one.
The equations are as follows.
\eqn\mixed{\eqalign{&a_1=zb_1\qquad   b_2=za_2\cr &b_1 b_2-c_1 c_2=\nu\cr}}
The base of this CY fourfold is a
noncontractible five-cycle containing an ${\bf S}^2$ and an ${\bf S}^3$.
We can visualize the base of the manifold in the following diagram where 
we take $\nu$ real and positive.  The
horizontal edges are ${\bf S}^2\times {\bf S}^1$ with the ${\bf S}^1$
contracting at the opposite edge while the angled edges are ${\bf S}^3$.
Since the minimal ${\bf S}^2$ is isolated at $b_1=0$ inside the five-cycle,
the ${\bf S}^2\times {\bf S}^1$ three-cycle, although homologous
to the ${\bf S}^2$ at $c_1=0$, is not connected to it by a flat direction
that preserves the radius of the ${\bf S}^2$.  This cycle is not 
supersymmetric since it is not Lagrangian.

\vskip .5cm
%Diagram 4
\centerline{\epsfxsize=0.5\hsize\epsfbox{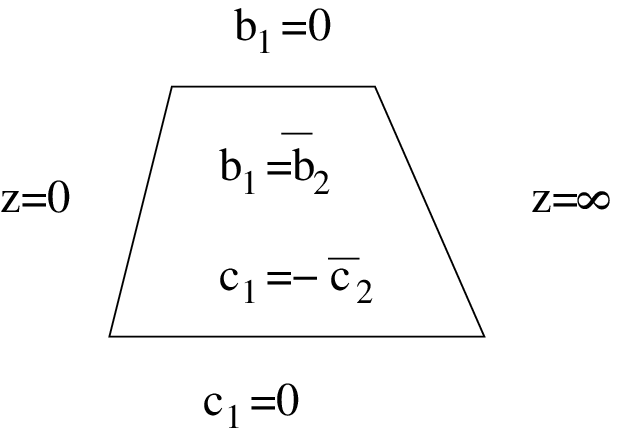}}
\bigskip
\centerline{\vbox{\noindent{\bf Diag. 4.}
Base of the Mixed Resolution}}

\noindent The normal directions are $b_1-{\bar b}_2$ and $c_1+{\bar c}_2$
constrained by $b_1 b_2-c_1 c_2=\nu$.  There is a zero mode direction for
the ${\bf S}^2$ along $b_1=b_2=0$ and for ${\bf S}^3$ inside the 
five-cycle.  In this 
phase the modes associated with the deformation $\nu$ are massive
because the four-cycle of topology ${\bf S}^3\times {\bf S}^1$
shrinks to an ${\bf S}^3$ at the poles
of the ${\bf S}^2$.

\subsec{Geometric Transitions}

To show that there are possible transitions between all of the phases that are
described above, we note that there is a common set of coordinates
for the small resolution, the nongeneric large resolution, and the
mixed resolution in which the holomorphic four-form is expressed as
\eqn\fourform{\Omega={da_1\wedge db_1\wedge db_2\wedge dc_2\over a_1 c_2} .}
The transversality of the intersections for the various resolutions 
implies that $\Omega$ has no zeroes or poles.  The large resolution
generic $\Omega$ takes another form in general but can be continuously
deformed to the above $\Omega$ without encountering zeroes or poles.  
The existence of $\Omega$ ensures that the holonomy is no larger group
than $SU(4)$.   The mixed resolution can be described as a
manifold with base ${\cal O}(-1)+{\cal O}(-1)\rightarrow{\bf P}^1$ and 
fiber $T^*{\bf S}^1$ where the fiber degenerates along $b_1 b_2=\nu$
(see \mixed ).  The large resolution can be described as a $T^*{\bf S}^1$
fibration over $T^*{\bf S}^3$ that degenerates
along $b_1=0$ and $b_2=0$.  The base in each case has a nonvanishing
and covariantly constant holomorphic three-form.  The pullback of this form
to the CY fourfold must have a zero or pole.  Otherwise, we could
obtain a covariantly constant one-form by wedging with the 
antiholomorphic four-form and taking the Hodge star dual.  This is
impossible since the manifold is simply connected.  The holonomy in all
cases is, thus, $SU(4)$.  The
dynamics of the possible transitions will be discussed in the next
section.

\newsec{Effective Theory of the Triple Intersection}

\subsec{The Small Resolution}

The effective two-dimensional theory for the triple intersection of
fivebranes in type IIA has a chiral $(0,4)$ supersymmetry.  Similarly,
type IIB on a CY fourfold has $(0,4)$ supersymmetry.  We require that the
spatial dimension of the two-dimensional theory be a circle so
that fivebrane deformations in the fivebrane theory are normalizable
and to avoid supergravity anomalies in the CY fourfold theory as 
discussed below.  To determine the 
zero modes of the triple intersection, we first consider the case of a smooth 
fivebrane wrapping a four-cycle on a six torus as discussed in \cdm .
This analysis is mostly inapplicable when the 
six-torus is decompactified, and the intersection is localized.  We do not
have a rigorous derivation of the zero mode spectrum but argue that the
self-intersecting fivebrane can be resolved to a smooth fivebrane
wrapped on a ${\bf P}^2$ face of the manifold ${\bf P}^3$.
Requiring that the first Chern class
of the total space vanish, we obtain ${\cal O}(-4)
\rightarrow{\bf P}^3$.  This resolution reduces the supersymmetry to 
$(0,2)$.  The moduli
space of ${\bf P}^2$'s inside ${\bf P}^3$ is ${\bf P}^3$.
We illustrate the fivebrane
resolution of the singularity as a tetrahedron in the following diagram.  
The fivebrane is
wrapped on the shaded face.

\vskip .5cm
%Diagram 5
\centerline{\epsfxsize=0.5\hsize\epsfbox{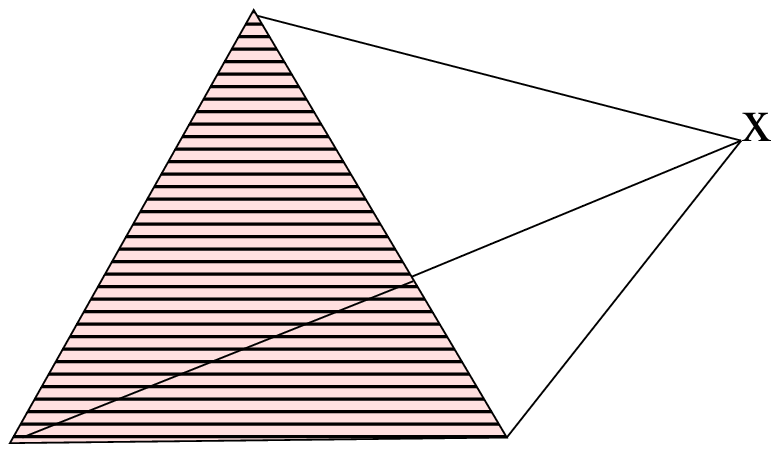}}
\bigskip
\centerline{\vbox{\noindent{\bf Diag. 5.}
Small Fivebrane Resolution-Toric Diagram of ${\bf P}^3$}}

If we exclude deformations of the fivebrane that intersect the point
labeled $x$, the moduli space of deformations is ${\bf C}^3$.  When the
fivebrane reaches $x$, there is a ``flop'' transition but no singularity.
The moduli space of the fivebrane inside ${\bf P}^3$ encompasses all
of the flopped phases of the CY fourfold with no singularities.
If we were to remove a
${\bf P}^2$ face on which the fivebrane is wrapped and wrap a fivebrane
on the three other faces, we would return to the singularity of the
triply intersecting fivebrane in ${\bf C}^3$.  The resolution is a
compactification in which the fivebrane flops to the compactifying
face.  Since this manifold allows for fivebrane instantons, it is 
natural to conjecture that this transition is mediated by a fivebrane
instanton in ${\bf P}^3$.  The transition is depicted in the following 
diagram.

\vskip .5cm
%Diagram 6
\centerline{\epsfxsize=1.0\hsize\epsfbox{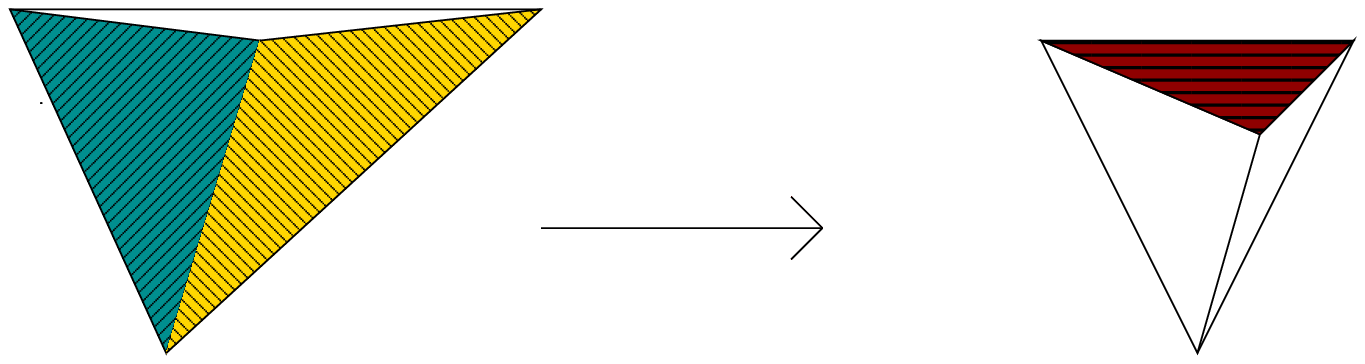}}
\bigskip
\centerline{\vbox{\noindent{\bf Diag. 6.}
Instanton Transition from Triple Intersection to Smooth
Fivebrane}}

Additional arguments for selecting the above manifold are that
the broken translational invariance due
to the intersection provides six real modes on ${\bf R}^6$ and one
translational mode for the compactified dimension of 
eleven-dimensional supergravity.  Translations on the complex
plane acquire a mass because of the conical deficit.  As we have seen in the 
supergravity analysis, the manifold with the fivebrane is not Kahler so
there is no scale modulus, and the ${\bf P}^3$ cannot shrink.  From the
supergravity equations we also see that the scale of the ${\bf P}^3$ should
be set by the number of fivebranes which in our case is one.
The generator of the second homology class inside the ${\bf P}^2$ 
of the fivebrane
has genus zero and positive self-intersection and is 
therefore self-dual.  All together
we have the following worldsheet theory for scalar bosons (B) and 
fermions (F)
\eqn\worldsheetsmall{\eqalign{N_L^B&=7 \qquad N_R^B=8\cr N_L^F&=0 
\qquad N_R^F=8\cr c_L&=7 \qquad c_R=12\cr}}
where $R$ is for right movers and $L$ for left movers 
and $c$ is the central charge.

Now compare this result with the CY fourfold theory in the ${\cal O}(-1,-1)+
{\cal O}(-1,-1)\rightarrow {\bf P}^1\times{\bf P}^1$ phase.  In two
dimensions all moduli of the vacua must be dynamical variables.
Supersymmetry
constrains the number of left moving matter scalars to be a multiple of four.
Reduction 
of the Kahler form and the two $B$ fields on the ${\bf P}^1$'s 
yields six nonchiral scalars.  The
Ramond-Ramond (RR) scalar provides another nonchiral scalar.  
Additionally, there is one antiself-dual
four-cycle which provides a left moving scalar.  Since the four-cycle is
isolated, its self-intersection must be negative.  Supersymmetry pairs the
left movers with fermions.  In the compact case the number of right moving
fermions is determined by index theorems to be proportional to the
number of three-cycles \dm .  We assume that if the small resolution
is a limit of a compact CY fourfold with three-cycles that these modes
are nonnormalizable in this limit.  The dilaton goes
into the supergravity multiplet which includes four right moving
gravitinos and four left moving spin one-half fermions.  
In order to match the supersymmetry of the 
effective theory of the fivebrane resolution which is broken to
$(0,2)$ by the CY fourfold, we need to wrap a Dirichlet fivebrane around the
four-cycle. The fivebrane also adds a $U(1)$ field to match the $U(1)$
from the
membrane potential reduced on the ${\bf P}^1$ of ${\bf P}^2$.  
The moduli of the four-cycle (Kahler parameters) are zero modes 
in the two-dimensional effective theory of the fivebrane.  Additionally, the
RR fields couple to the fivebrane giving the same zero modes as for the 
CY fourfold.  The scalar from the antiself-dual four-cycle is 
charged under the 
$U(1)$ of the fivebrane.  
The zero mode
spectrum of the matter multiplets agrees with that found for the triple
intersection except that left and right movers are exchanged.  
These zero modes would be precisely those of the 
heterotic string on a circle if seventeen
right movers were not missing.  

The conformal anomaly of this theory is the reduction of the fivebrane 
anomaly which is cancelled by bulk counterterms as has been shown
in great detail by many authors.  On the other hand type IIB on a
compact CY fourfold is not anomalous as shown in \dm .  Accordingly,
the appropriate counterterms are not present for type IIB on a
compact CY fourfold.  The T-duality relation ensures that these
terms are present to cancel the anomaly for our noncompact case.
From the discussion we see that a compact CY fourfold cannot be constructed
easily out of intersecting fivebranes because the conformal anomaly of the 
$(0,4)$ theory generally requires a bulk counterterm.  The supergravity
multiplet in type IIA is not anomalous, but there is a two-dimensional anomaly
for this multiplet in type IIB theory.  The T-dual counterterms will not
cancel this anomaly.  Consistency requires that either the supergravity
multiplet be decoupled or that there are further corrections to type IIB
on a noncompact CY fourfold.  As we mentioned above the two-dimensional
theory is compactified on a circle so there is no supergravity anomaly.
Threebrane instantons may smooth the flop
transitions.

Note that the case of four triply intersecting fivebranes can be resolved
by fivebrane instantons to a smooth fivebrane wrapped around a $K3$
surface embedded in ${\bf P}^3$.  Counting deformations and two-cycles
we obtain $72$ right moving scalars and fermions and $88$ left
movers.  The amount of supersymmetry makes this theory likely
equivalent to the heterotic string on the product of a two-torus
and a noncompact CY threefold.

{\it{Local Mirror Symmetry}}

The resolution of the triple intersection singularity by a fivebrane
instanton reduces the supersymmetry to $(0,2)$.  There is another 
resolution that preserves $(0,4)$ supersymmetry and should be 
equivalent to the CY fourfold in the small resolution phase by local
mirror symmetry and T-duality.  Our discussion here has many gaps as
we are not able to fully analyze this system.  Despite these gaps this
presentation may be useful for further analysis of this system.
The techniques of local mirror
symmetry applied to a gauged linear sigma model are discussed in \kmvhi .
Starting with the linear sigma model \charge , the mirror 
CY fourfold is obtained from the following equations.
\eqn\mirror{\eqalign{z&=1+e^u +e^v +e^w +e^{v-u-t_1}+e^{v-w-t_2}\cr
z&=xy\cr}}
The first equation is a noncompact holomorphic surface 
fibered over the $z$ plane
that degenerates at two points in the $z$ plane for finite $u$, $v$,
and $w$ and at $z=1$ for $u, v, w\rightarrow -\infty$.  The second equation is
a ${\bf C}^*$ fibration that degenerates at $z=0$.  The total space has
no singularities for generic $t_1$ and $t_2$.  At each 
degeneration of the holomorphic surface
for finite $u$, $v$,
and $w$, we expect but cannot prove that an 
${\bf S}^2$ shrinks.  We can form two
four-cycles of topology ${\bf S}^4$ by connecting $z=0$ with the two
points where the holomorphic surface degenerates.  A linear combination 
of these two cycles as well as a cycle joining $z=0$ and $z=1$ is
expected to be mirror to the four-cycle, ${\bf P}^1\times {\bf P}^1$, 
and antiself-dual.  Mirror symmetry requires that this four-cycle be 
rigid as there should not be a compact two-cycle.  It is not clear to
me how this works.  The mirror of the 
zero-cycle is expected to be noncompact.

The two ${\bf S}^4$'s are combinations of cycles with Hodge type $(1,3)$
and $(3,1)$, and their deformations correspond to complex structure 
deformations.  Generically, the deformations of middle-dimensional
Lagrangian cycles are not normalizable but are
logarithmically divergent.  In two dimensions massless scalars have
logarithmically divergent two-point functions so that this type of 
divergence is acceptable here.  Each ${\bf S}^4$ yields three
nonchiral scalars in the effective two-dimensional theory 
from the complex structure deformations and the   
reduction of the RR four-form.   The $(2,2)$ cycle mirror to
${\bf P}^1\times {\bf P}^1$ gives a
left moving scalar, and the RR scalar gives a nonchiral scalar. 
Supersymmetry generates left moving fermionic partners.  With a lot of
assumptions, the zero mode
spectrum is the same as what we previously found.

By T-dualizing on the circle of the ${\bf C}^*$ fibration, we obtain a type 
IIA fivebrane wrapped on the noncompact surface at $z=0$.
\eqn\fivesmall{1+e^u +e^v +e^w +e^{v-u-t_1}+e^{v-w-t_2}=0}
Introducing another coordinate and taking the coordinates to be projective,
we can write this equation as 
\eqn\proj{xyz+y^2z+wyz+yz^2+\lambda_1 wxz+\lambda_2 wxy=0.}
If either or both $\lambda_1$ and $\lambda_2$ diverge, the singularity
of the triply intersecting fivebrane is recovered.  Otherwise, there
are two isolated singularities when three of the coordinates are zero
which are on the boundary of \fivesmall .  The boundary is the intersection
of $wxyz=0$ and \proj .  We will not in this paper be able to determine
the deformations and two-cycles of the noncompact four-cycle to analyze
the effective two dimensional theory.

\subsec{The Mixed Resolution}

The phase where we separate one fivebrane from the other two
corresponds to the mixed resolution.  The effective theory at the 
intersection of two fivebranes is four-dimensional, and the matter 
content in this phase
is nonchiral.  There are four nonchiral translational modes moving
the intersection of the two fivebranes in ${\bf R}^4$
as well as a supersymmetric completion of
fermions.  The supersymmetry breaking by the third fivebrane occurs
at a distance from the intersection and only affects fields that interact
with the bulk.  The zero modes restricted to the intersection are
unaffected by this breaking and are nonchiral.  Going to the CY fourfold, 
we argue that the effective theory is four-dimensional and nonchiral.
The ${\bf S}^2$ has four flat, normal spacetime directions, 
$T^*{\bf S}^1\times{\bf S}^1\times{\bf R}$.  The $T^*{\bf S}^1$ fibration
degenerates, and the supersymmetry is broken in half at a distance of
order $|\nu|^{1\over 2}$ from the ${\bf S}^2$.  Zero modes which come from
reducing the theory on ${\bf S}^2$ are not affected by this breaking since 
they do not interact with the bulk.
We obtain a
left-right symmetric combination of four scalar bosons and four
fermions as well as a $U(1)$ gauge field 
from the ${\bf S}^3$.  The zero mode scalars come from
reducing the Kahler form, two $B$ fields, and RR four-form on ${\bf S}^2$.
In the four dimensions of spacetime normal to the ${\bf S}^2$ 
($T^*{\bf S}^1\times{\bf S}^1\times{\bf R}$), the reduced RR four-form 
is dual to a scalar.  For the effective four-dimensional theory we
expect these modes to be nondynamical.  
If we ``compactified'' this theory on $T^*{\bf S}^1$, we
would have a two-dimensional theory with the following zero mode spectrum.
\eqn\worldsheetmix{\eqalign{N_L^B&=4 \qquad N_R^B=4\cr N_L^F&=4 
\qquad N_R^F=4\cr c_L&=6 \qquad c_R=6\cr}}
We would actually have a continuous family of these theories 
labeled by $\nu$ which was what we found before we realized that 
the effective theories are four-dimensional.

The modes related to the complex parameter $\nu$ are massive because of
the four-cycle that shrinks at the poles of the ${\bf S}^2$.  Unless there
is a five-form flux on the five-cycle, the four-form gauge field is
pure gauge.  The situation here is analogous to the magnetic field for
a Dirac monopole outside of a two-sphere of radius $r$ in ${\bf R}^3$.
If there is a magnetic field, its flux is quantized.  The total energy
of the field outside of the two-sphere is $g^2\over r$ where $g$ is
the magnetic charge unit.  Our case is more complicated because there
are two scales, the areas of the ${\bf S}^2$ and the ${\bf S}^3$, and
there are three normal directions.  One of these directions is compact.
There is also the zero mode direction for the minimal ${\bf S}^2$ which
potentially causes a divergence in the energy.  The minimal ${\bf S}^2$
is a set of measure zero in the five-cycle, and to estimate the energy
we assume there is no divergence.   A rough estimate of the 
energy of the five-form field
is to divide the volume of a minimal eight-ball surrounding the five-cycle
by the area squared of the five-cycle.  Without knowing the metric and
taking the square of the radius to be the mean square radius of 
the ${\bf S}^2$ 
and the ${\bf S}^3$, $|\nu|^{1\over 2}R$, we obtain an energy scaling as
$1\over |\nu|$.  
Assuming that this energy should be independent of the Kahler parameter
uniquely determines that the energy scale as $1\over |\nu|$.  The zero
modes of the Kahler parameter, the expectation value of the RR $B$ field
on the ${\bf S}^2$, and that of the RR four-form form a massive 
supermultiplet 
in the presence of five-form
flux.  Note that the mass diverges at the origin of the mixed
resolution so that dynamical transitions into this phase when the 
${\bf S}^3$ shrinks are not possible.
On the fivebrane side this flux should correspond to a magnetic field.
Another argument to support the above hand waving is the following.  If
$\nu\rightarrow\infty$ what remains is the small resolution of the
conifold with no flux, and this hypermultiplet should be massless.  On the
other hand, taking the Kahler parameter to infinity leaves the ${\bf S}^3$
resolution of the conifold, and this hypermultiplet should be massive.
There is another massive supermultiplet with mass 
proportional to $|\nu|^{3/2}$ from the threebrane wrapping ${\bf S}^3$
as in the conifold \s .  The four massive modes are the ${\bf S}^2\times 
{\bf S}^1$ directions and the global $U(1)$ charge mode.  

\subsec{The Large Resolution}

The phase where all three fivebranes are separated corresponds to the 
large resolution.  There are no normalizable translational zero modes 
in this phase. The effective theory for two fivebranes joined
by a fourbrane is a four-dimensional theory with a decoupled $U(1)$
parametrized by the separation of the fivebranes in ten-dimensional
spacetime and their difference in position in the compactified
eleventh dimension.  The
two independent separations of the fivebranes and differences in 
positions along the
compactified eleventh dimension will not be normalizable modes.  One 
way to see this is that these modes correspond to complex parameter
deformations of
${\bf S}^3$ inside $T^*{\bf S}^3$ which are logarithmically
divergent.

On the CY fourfold side we will explain why the theory is trivial
at low energies.  There is a
subtlety concerning the deformations of the four-cycles.  The four-cycles
are topologically ${\bf S}^3\times{\bf S}^1$'s, and the parameter
governing the size of the ${\bf S}^3$ is distinct from that determining 
the size of ${\bf S}^1$.  Deforming an ${\bf S}^3$ inside $T^*{\bf S}^3$
would yield a logarithmically divergent four-dimensional mode and
integrating over the zero mode $T^*{\bf S}^1$ direction would generate 
an additional divergence for a two-dimensional mode.
We obtain no normalizable scalar modes from the four-cycles in this phase.  To
understand this phase we need to remember that each ${\bf S}^3$ has a
cylindrical zero mode direction along $T^*{\bf S}^1$.  In the 
four dimensions of spacetime ${\bf S}^1\times{\bf R}\times
T^*{\bf S}^1$ remaining after reduction on ${\bf S}^3$, 
the RR scalar, the two $B$ fields, and the dilaton
are on an equal footing since scalars and two-forms are dual.  
Again, one would expect these modes to be nondynamical in four
dimensions.  The same argument about supersymmetry breaking
in the mixed resolution applies here.  We should obtain no chiral
fermions in this phase.  
``Compactifying'' the theory on $T^*{\bf S}^1$ would yield a trivial
two-dimensional theory at low energies with no conformal anomaly.
We summarize this below.  
\eqn\worldsheetlarge{\eqalign{N_L^B&=0 \qquad N_R^B=0\cr N_L^F&=0 
\qquad N_R^F=0\cr c_L&=0 \qquad c_R=0\cr}}
As in the mixed resolution the puzzle that we appear to have a family of
two-dimensional vacua parametrized by $\mu$ and $\nu$ is resolved by
realizing that the effective theory is really four-dimensional here.
There are two massive supermultiplets from threebranes wrapping the 
independent ${\bf S}^3$'s.  The zero modes are the global $U(1)$ and
$T^*{\bf S}^1\times {\bf S}^1$.  These supermultiplets become massless
at the singularity.  

The theories we have found are not strictly two-dimensional because of
the bulk counterterms needed for anomaly cancellation and the 
zero mode directions.  The mixed and large resolutions are effectively
four-dimensional and nonchiral.  Any dynamical transition from
the small resolution into the other phases seems unlikely.  Wrapping a
large number of fivebranes on ${\bf P}^1\times{\bf P}^1$ would yield
a large $N$ two-dimensional gauge theory without dynamics.
On the other hand the mixed and large resolutions are analogous to the 
${\bf S}^2$ and ${\bf S}^3$ resolutions of the conifold.  It
is natural to conjecture that there is a duality relating fivebranes
wrapped on ${\bf S}^2\times T^*{\bf S}^1\times{\bf S}^1\times {\bf R}$
in the mixed resolution with threeform flux on one of the ${\bf S}^3$'s
in the large resolution since the effective theories are 
four-dimensional.  By taking the radius of the other ${\bf S}^3$
to infinity, one retrieves the two resolutions of the conifold.
It might be interesting to explore further the equations
for the fivebrane following from local mirror symmetry to have a
purely geometric understanding of the fivebrane manifold in the 
mixed and large phases.  Determining whether there is a compact
manifold with these transitions would be interesting.  Also, can
one understand a triple intersection of fivebranes in type IIB
similarly as giving an effective two-dimensional theory of two
$(2,2)$ supersymmetric vector multiplets in type IIA?
The triple intersection theory is related to string theory black holes.
In type IIA the number of intersecting fivebranes is limited by the conical
deficit angle, but the number in eleven dimensional supergravity is
unlimited.  The various phases should still be present as the eleventh
dimension opens up.  There are many additional questions including 
nonperturbative terms in these theories left for further work.

\bigskip\centerline{\bf Acknowledgments}\nobreak
I wish to thank J. Distler, W. Fischler, T. Husain, and A. Iqbal for many
helpful discussions.  This material is based upon work supported in part
by the National Science Foundation under Grant No. 0071512.

\listrefs

\end